\documentclass[aps,prl,reprint,preprintnumbers,showpacs,showkeys,superscriptaddress]{revtex4-1}
\usepackage{amsmath}
\usepackage{bm}
\usepackage{hyperref}
\usepackage{mathrsfs}
\usepackage{times}
\usepackage{amsfonts}

\newcommand\vek[1]{\bm{#1}}
\newcommand\bra[1]{\langle #1|}
\newcommand\ket[1]{|#1\rangle}
\newcommand\gr[1]{\mathrm{#1}}

\begin{document}

\title{Massive Nambu--Goldstone Bosons}

\author{Haruki Watanabe}
\email{hwatanabe@berkeley.edu}
\affiliation{Department of Physics, University of California, Berkeley, California 94720, USA}

\author{Tom\'{a}\v{s} Brauner}
\email{tbrauner@physik.uni-bielefeld.de}
\affiliation{Faculty of Physics, University of Bielefeld, 33615 Bielefeld, Germany} 
\affiliation{Department of Theoretical Physics, Nuclear Physics Institute ASCR, 25068 \v Re\v z, Czech Republic}

\author{Hitoshi Murayama}
\email{hitoshi@berkeley.edu}
\affiliation{Department of Physics, University of California, Berkeley, California 94720, USA}
\affiliation{Theoretical Physics Group, Lawrence Berkeley National Laboratory, Berkeley, California 94720, USA}
\affiliation{Kavli Institute for the Physics and Mathematics of the Universe (WPI), Todai Institutes for Advanced Study, University of Tokyo, Kashiwa 277-8583, Japan}

\begin{abstract}
Nicolis and Piazza have recently pointed out the existence of Nambu--Goldstone-like excitations in relativistic systems at finite density, whose gap is \emph{exactly} determined by the chemical potential and the symmetry algebra. We show that the phenomenon is much more general than anticipated and demonstrate the presence of such modes in a number of systems from (anti)ferromagnets in magnetic field to superfluid phases of quantum chromodynamics. Furthermore, we prove a counting rule for these massive Nambu--Goldstone bosons and construct a low-energy effective Lagrangian that captures their dynamics.
\begin{center}
\emph{Dedicated to Ji\v{r}\'{\i} Ho\v{s}ek on the occasion of his 70th birthday.}
\end{center}
\end{abstract}

\preprint{BI-TP 2013/02, IPMU13-0056}
\maketitle


\emph{Introduction.}---Trying to understand collective behavior of matter in nonlinear many-body systems is a challenge common to many areas of physics. At long distances and low temperatures, excitations with vanishing or small gap (mass) dominate the dynamics. The concept of spontaneous symmetry breaking has been crucial for its understanding, as it unambiguously predicts existence of gapless excitations---the Nambu--Goldstone bosons (NGBs)---such as phonons or magnons. For nearly five decades, however, their correct counting and dispersion relations eluded consistent understanding.  Recently, the present authors developed a unified framework to determine the number and dispersion relations of NGBs~\cite{Watanabe:2011ec,Watanabe:2012hr,*[see also ]Hidaka:2012ym}, including their redundancies~\cite{Watanabe:2013iia}.

Cases where exact statements can be made about gapped modes are rare though. Kohn's theorem states that a gas of charged particles with Galilean invariance, when exposed to a uniform magnetic field, sustains a collective mode with the cyclotron gap~\cite{Kohn}. Moreover, some soliton solutions to nonlinear equations saturate Bogomol'nyi-Prasad-Sommerfield bounds, allowing their energies to be determined based on symmetry alone~\cite{Bogomolny:1975de,*Prasad:1975kr}, albeit with limited applicability to observable systems. NGBs perturbed by explicit symmetry breaking effects acquire small gaps and are usually called pseudo-NGBs~\cite{Weinberg:1972fn}. Yet their gaps in general can be computed only approximately.  

Recently, Nicolis and Piazza~\cite{Nicolis:2012vf,*[see also ]Nicolis:2011pv} pointed out that the gaps of pseudo-NGBs can be determined in special circumstances. Considering Lorentz-invariant systems perturbed only by a chemical potential whose charge operator is spontaneously broken, they showed that the masses of some pseudo-NGBs can be computed exactly and are free of radiative corrections. We will call such states \emph{massive NGBs} (mNGBs). In the present Letter, we show that mNGBs appear in a much broader class of systems; the theory need not be Lorentz-invariant, or the chemical potential operator spontaneously broken. We provide a counting rule for the number of mNGBs and construct an effective Lagrangian description for them.


\emph{General argument.}---Consider a many-body system specified by the Hamiltonian $\mathcal{H}$ with an internal symmetry group $G$. In order to describe states with finite charge density, it is customary to introduce a chemical potential $\mu$ by  $\tilde{\mathcal{H}}\equiv \mathcal{H}-\mu Q$, where $Q$ is one of the generators of $G$.  The vacuum $\ket0$ is defined as the eigenstate of $\tilde{\mathcal{H}}$ with the lowest eigenvalue. Without loss of generality, we can take $\tilde{\mathcal{H}}\ket0=0$. Since the generators $Q_i$ of the Lie group $G$ commute with the Hamiltonian $\mathcal{H}$, they are all time-independent in the Heisenberg picture \emph{defined} by $\mathcal{H}$,
\begin{equation}
Q_i(t)\equiv\int d\vek x\,e^{i\mathcal{H}t-i\vek{P}\cdot\vek{x}}j_i^0(0)e^{-i\mathcal{H}t+i\vek{P}\cdot\vek{x}},
\label{charge}
\end{equation}
where $j_i^0(x)$ are the corresponding local charge densities.

When spontaneously broken, generators of the symmetry group $\tilde G$ of the full Hamiltonian $\tilde{\mathcal{H}}$ give rise to standard massless NGBs.  On the other hand, the observation made by Nicolis and Piazza guarantees existence of pseudo-NGBs, created by spontaneously broken generators that do not commute with $Q$, whose masses can be computed \emph{exactly} by group theory.

By the standard Cartan decomposition, explicitly broken generators can be split into pairs $Q_{\pm\sigma}$---the roots---such that 
\begin{equation}
[Q,Q_{\pm\sigma}]=\pm q_\sigma Q_{\pm\sigma},
\label{condition}
\end{equation}
where $Q_{\pm\sigma}$ are some complex linear combinations of explicitly broken generators and $(Q_{\pm\sigma})^\dagger=Q_{\mp\sigma}$. Let us now focus on the quantity $\lambda_{\sigma}\equiv\bra0[Q_{+\sigma}(t),j_{-\sigma}^0(0)]\ket0$, which is manifestly time-independent. Using Eq.~\eqref{charge}, inserting a complete set of eigenstates $\ket{n,\vek p}$ of momentum $\vek{P}$ and energy $\tilde {\mathcal{H}}$, and carrying out integration over space, we obtain
\begin{equation}
\begin{split}
\lambda_{\sigma}=&\sum_ne^{-i[E_n(\vek 0)-\mu q_\sigma]t}|\bra0j_{+\sigma}^0(0)\ket{n,\vek 0}|^2\\
&-\sum_ne^{i[E_n(\vek 0)+\mu q_\sigma]t}|\bra0j_{-\sigma}^0(0)\ket{n,\vek 0}|^2.
\end{split}
\label{commutator3}
\end{equation}
Provided that $\mu q_\sigma>0$, time-independence of the left-hand side and $E_n(\vek 0)\geq 0$ require $\bra0j_{-\sigma}^0(0)\ket{n,\vek 0}=0$ for each $n$. If $\lambda_\sigma$ is zero, implying $\bra0j_{+\sigma}^0(0)\ket{n,\vek 0}=0$ for each $n$ as well, $Q_{\sigma R}\equiv Q_{+\sigma}+Q_{-\sigma}$ and $Q_{\sigma I}\equiv-i(Q_{+\sigma}-Q_{-\sigma})$ cannot be spontaneously broken. Namely, there is no local field $\Phi(x)$ such that $\bra0[Q_{\sigma R,I}(t),\Phi(0)]\ket0\neq0$.

On the other hand, if $\lambda_{\sigma}\neq0$, $Q_{\sigma R}$ and $Q_{\sigma I}$ are broken spontaneously and there must be a state $\ket{n,\vek 0}$ with mass 
\begin{equation}
\tilde{\mathcal{H}}\ket{n,\vek 0}=E_n(\vek 0)\ket{n,\vek 0}=\mu q_\sigma\ket{n,\vek 0}
\label{eq:mass}
\end{equation}
such that $\bra0j_{+\sigma}^0(0)\ket{n,\vek 0}\neq0$ and $\bra0j_{-\sigma}^0(0)\ket{n,\vek 0}=0$. This is the mNGB associated with the pair $Q_{\pm\sigma}$.

Our derivation clarifies several points on mNGBs.  First, the assumptions of the underlying dynamics being Lorentz-invariant and $Q$ being spontaneously broken~\cite{Nicolis:2012vf} can clearly be dropped. Also, $\lambda_\sigma$ always plays the role of the order parameter for charges $Q_{\sigma R,I}$. Finally, $\bra0j_{-\sigma}^0(0)\ket{n,\vek 0}=0$ for all $n$ means $Q_{+\sigma}\ket0=0$, while if $\lambda_\sigma$ is nonzero, $Q_{-\sigma}\ket0\neq0$. This observation leads to a simpler, albeit less rigorous, understanding of mNGBs. Eq.~\eqref{condition} gives $[\tilde{\mathcal{H}},Q_{\pm\sigma}]=\mp\mu q_\sigma Q_{\pm\sigma}$ which implies that $Q_{-\sigma}\ket0$ has energy $\mu q_\sigma$. As there cannot be a state with energy lower than the vacuum, $Q_{+\sigma}\ket0$ has to vanish. Our argument is reminiscent of Kohn's theorem~\cite{Kohn}, allowing for a unified comprehension of the two phenomena.


\emph{Number of mNGBs.}---For a proper understanding of the low-energy dynamics of the system, it is important to know the number and dispersion relations of NGBs. Denoting the broken generators of $\tilde G$ as $\tilde Q_a$, the former is given by~\cite{Watanabe:2011ec,Watanabe:2012hr,*[see also ]Hidaka:2012ym},
\begin{equation}
n_\text{NGB}=n_\text{A}+n_\text{B},\quad
n_\text{A}=n_\text{BG}-\mathrm{rank}\,\tilde{\rho},\quad
n_\text{B}=\frac{1}{2}\mathrm{rank}\,\tilde{\rho},
\label{eq:nNGB}
\end{equation}
where $n_\text{BG}$ is the number of broken generators and 
\begin{equation}
\tilde{\rho}_{ab}\equiv-i\lim_{\Omega\to\infty}\frac1\Omega\bra0[\tilde{Q}_a,\tilde{Q}_b]\ket0,
\label{rho}
\end{equation}
$\Omega$ being the spatial volume. The type-A and B NGBs generally have linear and quadratic dispersions and correspond to type-I and II in the Nielsen-Chadha theorem~\cite{Nielsen:1975hm}, even though this is not always the case~\footnote{A type-B NGB is always of type-II, while a type-A one can be either type-I or type-II. Provided that there are no modes that are simultaneously type-A and type-II~\cite{Watanabe:2011ec,Watanabe:2013iia}, the two schemes coincide; the Nielsen-Chadha counting rule is then an equality.}. Each type-B NGB is described by a canonically conjugate pair of broken generators $\tilde{Q}_a$ and $\tilde{Q}_b$ with nonzero $\tilde{\rho}_{ab}$, hence two broken symmetries count as one degree of freedom, whereas type-A NGBs are stand-alone like in the original Nambu--Goldstone theorem.  

Here we address the question of counting the mNGBs~\footnote{The same issue was discussed briefly already in Ref.~\cite{Nicolis:2012vf}, concluding that ``we have \emph{one} gapped Goldstone mode for each \emph{pair} of broken NC generators.'' This statement, however, admits different interpretations, depending on the basis of generators.}. Namely, we show that their number is given by
\begin{equation}
n_\text{mNGB}=\frac12(\mathrm{rank}\,\rho-\mathrm{rank}\,\tilde{\rho}),
\label{counting}
\end{equation}
where the matrix $\rho$ is defined analogously to Eq.~\eqref{rho} for all generators of $G$ instead of just $\tilde{G}$. To that end, we have to further specify the structure of the Lie algebra. First, let us choose the maximal number of mutually commuting generators of $\tilde G$, including $Q$ itself, to form the Cartan subalgebra. By a proper choice of the vacuum $\ket0$, we can ensure that these Cartan generators are the only generators of $\tilde G$ that can have a nonzero vacuum expectation value~\cite{Watanabe:2011ec}. This alone does not prevent the explicitly broken generators from acquiring expectation values. Yet, $\pm\mu q_\sigma\bra0Q_{\pm\sigma}\ket0=\bra0[\mu Q,Q_{\pm\sigma}]\ket0=\bra0[Q_{\pm\sigma},\tilde{\mathcal{H}}]\ket0=0$ thanks to $\tilde{\mathcal{H}}\ket0=0$, so that $\bra0Q_{\pm\sigma}\ket0$ must vanish for any nonzero $q_\sigma$.  If we arrange the generators as $Q_i=(Q_{1R},Q_{1I},\dotsc,Q_{mR},Q_{mI},\tilde{Q}_1,\dotsc,\tilde{Q}_{\dim\tilde{G}})$, where $m\equiv(\dim G-\dim\tilde{G})/2$, the matrix $\rho$ becomes block-diagonal, $\rho=\mathrm{diag}(2i\tau_2\lambda_1,\dotsc,2i\tau_2 \lambda_m,\tilde\rho)$, $\tau_2$ being the second Pauli matrix. Thus, $\frac12(\mathrm{rank}\,\rho-\mathrm{rank}\,\tilde{\rho})$ counts the number of nonzero $\lambda_{\sigma}$'s. Assuming that there is at most one mNGB for each pair of $Q_{\pm\sigma}$, this proves our counting rule~\eqref{counting}. This assumption is natural if we can identify the mNGB state with $Q_{-\sigma}\ket0$ in a suitable large-volume limit. 

In the following, we provide examples of mNGBs, demonstrating the validity of Eq.~\eqref{counting} in physically interesting systems~\footnote{The simplest example is perhaps a free nonrelativistic particle.  It can be interpreted as a type-B NGB of a spontaneously broken centrally-extended $\gr{ISO(2)}$ symmetry~\cite{Brauner:2010wm}.  A nonzero (negative) chemical potential for the $\gr{SO(2)}$ generator---the operator of particle number---lifts its energy, making it a mNGB.  We then find $n_\text{mNGB}=1=\frac{2-0}2$ in accord with Eq.~\eqref{counting}.}.


\emph{Ferromagnet.}---The Hamiltonian of a ferromagnet enjoys the internal $G=\gr{O(3)}$ symmetry group of spin rotations. In the ground state, individual spins are aligned, breaking this symmetry down to its $\gr{O(2)}$ subgroup. The two broken generators give rise to a single type-B NGB with a quadratic dispersion relation at low momentum---the magnon~\cite{Watanabe:2012hr,Nielsen:1975hm}.

Consider now switching on a uniform magnetic field $\vek B$ oriented in the $z$-direction. This amounts to breaking the symmetry explicitly to $\tilde{G}=\gr{O(2)}$ by adding to the Hamiltonian $-\mu_m BS_z$ ($\mu_mB>0$), where $\vek S$ is the total spin operator and $\mu_m$ is the magnetic moment.  This term can be viewed as a chemical potential $\mu=\mu_m B$ for the generator $Q=S_z$.  Given that $[S_z,S_\pm]=\pm S_\pm$ where $S_\pm\equiv S_x\pm iS_y$, $S_-$ must excite a mNGB of gap $\mu$, which is just the magnon with energy lifted by the magnetic field~\cite{Leutwyler:1993gf}. The operator $S_+$ annihilates the ground state. Both of these assertions are easy to understand from the fact that the vacuum corresponds to the state with maximum spin in the direction of the magnetic field, and the magnon to an excitation caused by flipping one of the spins. Note that the counting rule~\eqref{counting} predicts the correct number of mNGBs, that is, $n_\text{mNGB}=\frac{2-0}{2}=1$. Also, the generator $Q$ in this example is not spontaneously broken, in contrast to the assumption made in Ref.~\cite{Nicolis:2012vf}.


\emph{Antiferromagnet.}---In the absence of a magnetic field, assume the spins are oriented alternately along the $z$-axis; $G=\gr{O(3)}$ is broken to $\gr{O(2)}$ just like in a ferromagnet. In this case there are two type-A NGBs, one for each broken generator.

Applying a magnetic field along the $z$-axis leads to an instability as the NGBs attempt to acquire masses $\pm\mu=\pm\mu_mB$. The ground state rearranges with alternating spins pointing in an orthogonal direction instead, say along the $x$-axis. Then $Q=S_z$ is a spontaneously broken generator which commutes with $\tilde{\mathcal{H}}$ and creates a gapless type-A NGB.  On the other hand, the pair of generators $S_x,S_y$ is explicitly broken, creating a mNGB with gap $\mu$. The magnetic field induces a small magnetization along the $z$-axis, and hence $\rho_{xy}=\bra0[S_x,S_y]\ket0\neq 0$. Consequently, $n_\text{mNGB}=\frac{2-0}{2}=1$, consistent with Eq.~\eqref{counting}. Such mNGBs have been discussed before in the context of the electron spin resonance phenomenon~\cite{Oshikawa}.


\emph{Relativistic Bose-Einstein condensation.}---As an explicit example where $\tilde{\rho}\neq0$, consider a theory of a complex scalar doublet $\phi$ with a global $\tilde{G}=\gr{SU(2)\times U(1)}$ symmetry,
\begin{equation}
\mathscr{L}=D_\mu\phi^\dagger D^\mu\phi-M^2\phi^\dagger\phi-\lambda(\phi^\dagger\phi)^2,
\label{sigmamodel}
\end{equation}
where $D_0\phi\equiv(\partial_0-i\mu)\phi$. This model features a relativistic Bose-Einstein condensation (BEC) phase for $\mu>M$, in which the symmetry is spontaneously broken to a $\gr{U(1)'}$ subgroup. The three broken generators produce two NGBs, one type-A and one type-B~\cite{Schafer:2001bq,*Miransky:2001tw}, consistent with Eq.~\eqref{eq:nNGB}, since one of the $\gr{SU(2)}$ charges develops nonzero density in the ground state, hence $\mathrm{rank}\,\tilde{\rho}=2$.

The type-B NGB in this model has an ``antiparticle'', carrying opposite charge. Its mass equals $2\mu$ and does not receive radiative corrections~\cite{Brauner:2006xm}. To see why, note that when $\mu=0$, the Lagrangian enjoys an extended internal symmetry, $G=\gr{SO(4)\simeq SU(2)_L\times SU(2)_R}$. This is most easily seen by defining a $2\times 2$ matrix $\Phi = \left( \phi, i\tau_2 \phi^* \right)$, which transforms under $G$ as $\Phi\to U_\text{L}\Phi U^\dagger_\text{R}$. Denote the generators of $G$ as $\vec L$ and $\vec R$, respectively; they are both given by a half of the Pauli matrices. The $\gr{SU(2)}$ rotations of the doublet $\phi$ now correspond to $\gr{SU(2)_L}$; the $\gr{U(1)}$ phase transformations are generated by $2R_3$. The remaining two generators of $\gr{SU(2)_R}$ are explicitly broken by the chemical potential $\mu$. In the BEC phase, the condensate can be chosen as $\bra0\Phi\ket0\sim\openone$ so that they are also broken spontaneously. Since the $R_\pm$ generators of $\gr{SU(2)_R}$ satisfy the commutation relation $[2R_3,R_\pm]=\pm2R_\pm$, Eq.~\eqref{commutator3} implies that $R_-$ creates a mNGB with mass $2\mu$, in agreement with the explicit calculation. Indeed, $n_\text{mNGB}=\frac{4-2}{2}=1$.  This example obviously admits a generalization to a large class of relativistic linear sigma models with chemical potential~\cite{Brauner:2005di,Andersen:2006ys}, the key ingredient being an extended global symmetry when the chemical potential is set to zero.


\emph{QCD-like theories.}---Quantum ChromoDynamics (QCD) with two degenerate quark flavors possesses an approximate global $\gr{SU(2)_L\times SU(2)_R}$ chiral symmetry.  A nonzero quark mass breaks this explicitly to the $G=\gr{SU(2)_V}$ subgroup generated by $\vec V\equiv\vec R+\vec L$. The chiral condensate in the QCD vacuum breaks the symmetry spontaneously in the same way, resulting in three pseudo-NGBs in the spectrum: the pions.

Nonzero chemical potential, $\mu_\text{I}$, for $V_3$ breaks the exact symmetry $G$ further to its $\tilde G=\gr{U(1)_I}$ subgroup, generated by $V_3$. While the mass of the neutral pion is insensitive to $\mu_\text{I}$, the masses of the charged pions become $m_\pi\pm\mu_\text{I}$. Consequently, once $\mu_\text{I}>m_\pi$, the positively charged pion undergoes BEC, breaking the residual $\tilde G$ symmetry spontaneously~\cite{Son:2000xc}. Therefore, the spectrum in the pion BEC phase exhibits one true, type-A NGB. However, the ground state has a nonzero isospin density, $\bra0V_3\ket0=-i\bra0[V_1,V_2]\ket0$, and Eq.~\eqref{counting} implies that there is also one mNGB. Such a state has indeed been found using effective field theory~\cite{Kogut:2001id} as well as various model approaches~\cite{He:2005nk,Andersen:2006ys} and can be identified with the neutral pion in the superfluid medium. As opposed to these approximate calculations, the result of Ref.~\cite{Nicolis:2012vf} nevertheless ensures that its mass is exactly equal to $\mu_\text{I}$. This follows from the commutation relation $[V_3,V_\pm]=\pm V_\pm$.

In the limit of massless quarks, the full symmetry becomes $G=\gr{SU(2)_L\times SU(2)_R}$; isospin chemical potential breaks this explicitly to $\tilde G=\gr{U(1)_L\times U(1)_R}$. Pion condensate now develops at any nonzero chemical potential, breaking $\tilde G$ spontaneously to $\gr{U(1)}$. Thus, there is one type-A NGB in the spectrum. Moreover, given the commutators $[V_3,R_\pm]=\pm R_\pm$ and $[V_3,L_\pm]=\pm L_\pm$, we find that $\bra0V_3\ket0=-2i\bra0[R_1,R_2]\ket0=-2i\bra0[L_1,L_2]\ket0\neq0$, as a result of which there are $\frac{4-0}2=2$ mNGBs according to Eq.~\eqref{counting}. This is consistent with explicit calculations; the additional mNGB has the quantum numbers of the $\sigma$ meson.

The presence of mNGBs has also been noted in the diquark BEC phase of two-color QCD. In case of two quark flavors, these are the three pions, with the mass equal to the baryon chemical potential, as observed in analytic calculations~\cite{Kogut:2000ek,Ratti:2004ra,*Strodthoff:2011tz} as well as on the lattice~\cite{Hands:2000ei}. An additional mNGB again appears in the limit of massless quarks. Similar conclusions can be reached for an arbitrary even number of flavors~\cite{Kogut:2000ek}.


\emph{Effective Lagrangian formalism.}---The effects of the chemical potential can be captured by a low-energy effective field theory (EFT). Assume first that at $\mu=0$, the symmetry group $G$ is broken spontaneously to its subgroup $H$. Insofar as $\mu$ is much smaller than the scale of this breaking, it can be treated as a perturbation. One constructs an EFT based on the coset space $G/H$~\cite{Leutwyler:1993gf,Leutwyler:1993iq} and introduces $\mu$ as a constant temporal gauge field~\cite{Kapusta:1981aa,*Haber:1981ts}; no additional free parameters are involved. Assuming spatial translational and rotational invariance, the lowest-order terms in the effective Lagrangian read~\cite{Leutwyler:1993gf}
\begin{equation}
\begin{split}
\mathscr{L}_\text{eff}=&c_a(\pi)\dot{\pi}^a+e_i(\pi)\mu^i+\frac12\bar{g}_{ab}(\pi)D_t\pi^a D_t\pi^b\\
&-\frac12g_{ab}(\pi)\vek{\nabla}\pi^a\cdot\vek{\nabla}\pi^b.
\label{efflag}
\end{split}
\end{equation}
Here $\pi^a$ ($a=1,\dotsc,\dim G/H$) are NG fields, while $g_{ab}(\pi)$ and $\bar{g}_{ab}(\pi)$ are both $G$-invariant metrics on the coset. Under an infinitesimal symmetry transformation defined by a set of parameters $\epsilon^i$ ($i=1,\dotsc,\dim G$), the coset fields change as $\delta\pi^a=\epsilon^ih^a_i(\pi)$, where $h^a_i(\pi)$ are the Killing vectors of the metrics. The covariant derivative is $D_t\pi^a\equiv\dot\pi^a-\mu^ih^a_i(\pi)$.

Explicit expressions can be obtained using the formalism of Ref.~\cite{Coleman:1969sm,*Callan:1969sn}. Denoting now the broken group generators as $T_a$ and the unbroken generators as $T_\rho$, we represent the coset element by $U(\pi)\equiv e^{iT_a\pi^a}$ and define the Maurer-Cartan form as $\omega_a(\pi)=T_i\omega_a^i(\pi)\equiv-iU(\pi)^{-1}\frac{\partial}{\partial \pi^a}U(\pi)$. Then~\cite{toappear},
\begin{equation}
\begin{gathered}
g_{ab}(\pi)=g_{cd}(0)\omega_a^c(\pi)\omega_b^d(\pi),\quad
e_i(\pi)=\nu_i^{j}(\pi)e_j(0),\\
h^a_i(\pi)\omega^b_a(\pi)=\nu_i^b(\pi),\quad
c_a(\pi)=-\omega_a^i(\pi)e_i(0),
\end{gathered}
\label{effcouplings}
\end{equation}
where $\nu^j_i(\pi)$ is defined by $T_j\nu^j_i(\pi)\equiv U(\pi)^{-1}T_iU(\pi)$. For consistency with the $G$-invariance of the action, the effective couplings $e_i(0)$ and $g_{ab}(0)$ must satisfy $f_{i\rho}^je_j(0)=0$ and $f_{\rho a}^cg_{cb}(0)+f_{\rho b}^cg_{a c}(0)=0$, where $f_{ij}^k$ are the structure constants of $G$.  Similar expressions hold for $\bar{g}$. Using Eq.~\eqref{effcouplings}, the effective Lagrangian is now completely fixed by the values of $g_{ab}(0)$ and $\bar{g}_{ab}(0)$, encoding decay constants of the NGBs, and of $e_i(0)$, expressing charge densities in the ground state.

With the effective Lagrangian at hand, one first has to determine the ground state triggered by the chemical potential; the coset parameterization can always be chosen so that this lies at $\pi=0$. Upon expansion in powers of the coset fields, the effective Lagrangian can be used to calculate any observable order by order in the derivative expansion.

Let us first consider systems with $c_a\neq0$. For a consistent derivative expansion, energy has to be counted as momentum squared, hence the term with two time derivatives is subleading. The leading-order potential is thus merely $V(\pi)=-e_i(\pi)\mu^i$. The chemical potential forces the ground state to rearrange so that $e_i(0)$ is maximally aligned with $\mu^i$, as in the ferromagnet. Using the expressions $\omega^i_a=\delta^i_a-\frac12f_{ab}^i\pi^b+\dotsb$ and $\nu^j_i=\delta^j_i-f^j_{ia}\pi^a+\frac12f^k_{ia}f^j_{kb}\pi^a\pi^b+\dotsb$, we obtain the precise condition that $\pi=0$ is a (local) minimum of the potential, $f^i_{ab}\mu^be_i(0)=0$, and the expansion of the Lagrangian to second order in the fields, $\mathscr L_\text{eff}=\frac12f^i_{ab}e_i(0)(\dot\pi^a\pi^b-\mu^jf^b_{jc}\pi^a\pi^c)$ plus terms with spatial derivatives. This is sufficient to assert the existence of a mNGB with mass given by Eq.~\eqref{eq:mass}. 

In the $c_a=0$ case, the vacuum and mass spectrum are determined by the term $\frac12\bar g_{ab}(\pi)D_t\pi^aD_t\pi^b$; energy now counts as the first power of momentum. The potential takes the form $V(\pi)=-\frac12\bar g_{ab}(0)v^a(\pi)v^b(\pi)$ where $v^j(\pi)\equiv\mu^i\nu^j_i(\pi)$. The ground state thus rearranges so that the chemical potential lies maximally in the subspace of broken generators, as in the antiferromagnet. Particularly simple expressions can be obtained when the coset $G/H$ is a symmetric space. Assuming that the chemical potential lies completely in the broken subspace [it is sufficient that $\bar g_{ab}(0)f^b_{c\rho}\mu^\rho=0$], the bilinear part of the effective Lagrangian becomes $\mathscr L_\text{eff}=\frac12\bar g_{ab}(0)(\dot\pi^a\dot\pi^b-\mu^a\mu^cf^\rho_{cd}f^b_{e\rho}\pi^d\pi^e)$ plus spatial derivative terms. This again leads to mNGBs in accord with the general argument.

Apart from the true NGBs and mNGBs, the EFT can predict pseudo-NGBs whose masses are \emph{not} given by Eq.~\eqref{eq:mass}~\footnote{A simple example is provided by $G/H=\gr{O(3)/\{\}}$ with zero charge densities in the ground state. Three NGBs result, two of which acquire a mass upon switching on a chemical potential for one of the $\gr{O(3)}$ generators. Yet only one of them is a mNGB, in accord with our counting rule~\eqref{counting}. The existence of such pseudo-NGBs was also noticed very recently in Ref.~\cite{Nicolis:2013sga}.}. For such modes, $\lim_{\vek p\to\vek0}\bra0j^0_{\pm\sigma}(0)\ket{n,\vek p}=0$ at fixed nonzero $\mu$, thus not contributing to $\lambda_\sigma$, Eq.~\eqref{commutator3}. In the limit $\mu\to0$, their masses vanish and they become true NGBs. Their number can be inferred from known counting rules, namely as the number of NGBs at $\mu=0$ minus the numbers of NGBs and mNGBs at nonzero $\mu$, given by Eqs.~\eqref{eq:nNGB} and \eqref{counting}.

When the chemical potential is large, perturbing $G/H$ is not adequate. One can then describe both NGBs and mNGBs by an EFT based on the $G/\tilde{H}$ coset space, $\tilde{H}$ being the unbroken subgroup of the ground state in presence of $\mu$. In this approach, effective couplings may implicitly depend on $\mu$. In terms of generators, $G/\tilde{G}$ is K\"ahler and symplectic, and hence all generators in $\mathfrak{g}/\tilde{\mathfrak{g}}$ can be paired in $\rho$, giving mNGBs. The rest of generators in $\tilde{\mathfrak{g}}/\tilde{\mathfrak{h}}$ represent true NGBs of either type. In general, there may be other light modes, not automatically captured by the EFT, whose masses are close to mNGBs~\footnote{For example, when the spontaneous symmetry breaking is triggered by the chemical potential itself, the associated Higgs mode has a gap comparable to $\mu$.}. Such modes have to be added to the EFT as matter fields~\cite{Coleman:1969sm,*Callan:1969sn}.

In any case, the EFT reproduces the predicted masses of mNGBs. Symmetry guarantees that the masses do not acquire any higher-order corrections. Of course, the utility of the EFT is not limited to the mass spectrum. The nonlinear structure of the Lagrangian~\eqref{efflag}, dictated by symmetry, allows one to make predictions for any other low-energy observable.


We would like to thank Giorgio Torrieri for pointing the paper~\cite{Nicolis:2012vf} out to us, Siddharth A.~Parameswaran for informing us about Kohn's theorem~\cite{Kohn}, and Masaki Oshikawa for telling us about the electron spin resonance~\cite{Oshikawa}. The fact that $c_a(\pi)$ can in some theories such as the model of Eq.~\eqref{sigmamodel} be expressed in terms of the Maurer-Cartan form was first pointed out to us by Huan-Hang Chi. We are grateful to Toru Kojo for inspiring discussions. The work of T.B.~was supported by the Sofja Kovalevskaja program of the Alexander von Humboldt Foundation. He further acknowledges the hospitality of the Institute for the Physics and Mathematics of the Universe, where the presented work was initiated.  The work of H.M.~was supported in part by the U.S.~DOE under Contract DE-AC03-76SF00098, in part by the NSF under grant PHY-1002399, the JSPS grant (C) 23540289, and in part by WPI, MEXT, Japan. H.W.~appreciates financial support from Honjo International Scholarship Foundation.


\bibliography{references}

\end{document}